\documentclass[a4paper,conference]{IEEEtran}
\usepackage[left=1.62cm,right=1.62cm,top=1.9cm]{geometry}

\usepackage{pgffor}
\usepackage{xprintlen}
\usepackage{booktabs}
\usepackage{amsmath,amssymb}
\usepackage{graphicx}
\usepackage{xspace}
\usepackage{color}
\usepackage{epstopdf}
\usepackage{algorithm}
\usepackage{algorithmic}
\usepackage{booktabs}
\usepackage{lipsum}
\usepackage{caption}
\usepackage{subcaption}
\usepackage{setspace}
\definecolor{azure}{rgb}{0.0, 0.5, 1.0}
\usepackage{pgfplots, tikzscale}
\usetikzlibrary{plotmarks}
\usetikzlibrary{plotmarks,patterns}
\pgfplotsset{compat=newest}
\usepackage{comment}
\usepackage[inline]{enumitem}
\usepgfplotslibrary{external}
\usepackage{fixltx2e}
\usepackage{stfloats}
\usepackage{float}
\floatstyle{plain}
\restylefloat{table}
\usepackage{tabularx}
\usepackage{caption} 
\usepackage[normalem]{ulem}

\graphicspath{ {figures/} }

\title{\fontsize{18.9}{18.9}\selectfont NL-COMM: Enhanced Video Streaming via Advanced Non-Linear Processing  }

\author{\IEEEauthorblockN{Marcin Filo, George N. Katsaros, Chathura Jayawardena, and Konstantinos Nikitopoulos}
\IEEEauthorblockA{Wireless Systems Lab, 5G \& 6G Innovation Centre, \\Institute for Communication Systems, University of Surrey, Guildford GU2 7XH, UK}
}

\begin{document}
\bstctlcite{IEEEexample:BSTcontrol}
\maketitle

\begin{abstract}
With video streaming now accounting for the majority of internet traffic, wireless networks face increasing demands, especially in densely populated areas where limited spectral resources are shared among many devices. While multi-user (MU)-MIMO technology aims to improve spectral efficiency by enabling concurrent transmissions over the same frequency and time resources, traditional linear processing methods fall short of fully utilizing available channel capacity. These methods require a substantial number of antennas and RF chains, to support a much smaller number of MIMO streams, leading to increased power consumption and operational costs, even when the supported streams are of low rate. 
In this demo, we present NL-COMM, an advanced non-linear MIMO processing framework, demonstrated for the first time with commercial off-the-shelf (COTS) user equipment (UEs) in a fully 3GPP-compliant environment.
In addition, also for the first time, the audience will compare and assess the quality of live, over-the-air video transmission from four concurrently transmitting UE devices, alternating between current state-of-the-art MIMO detection algorithms and NL-COMM. Key gains of NL-COMM include improved stream quality, halving the number of required base station antennas without compromising stream quality compared to linear approaches, as well as achieving antenna overloading factors of 400\%.

\end{abstract}


\section{Introduction}
Video streaming transmissions now constitute more than 73\% of overall internet traffic \cite{ericsson2024mobile}, placing substantial demands on wireless infrastructure. 
Despite often high-SNR conditions at the receiver side, users frequently encounter issues like unstable or delayed video calls when transmitting in densely populated areas. These challenges arise from the limited spatial, frequency, and temporal resources that must be shared among multiple devices in congested environments.
To mitigate these challenges and improve spectral efficiency, wireless infrastructure
(both 3GPP and IEEE 802.11-based) commonly employs multiple-input, multiple-output (MIMO) technology, enabling multiple concurrent transmissions over the same frequency and time resources.

However, existing multi-user (MU)-MIMO deployments primarily rely on linear processing methods, such as Minimum Mean Square Error (MMSE) and Zero-Forcing (ZF), to demultiplex the interfering transmission streams. These methods work by decomposing the MIMO channel into separate single-user channels, making them computationally efficient and relatively straightforward to implement. Despite these computational advantages, linear processing methods significantly underutilize the available channel capacity.
As a result, and as we also show in this demo, linear processing requires a disproportionately large number of antennas and RF chains to support only a small number of streams. This leads to an unnecessary increase in power consumption and operational costs.

The inefficiencies of traditional linear MIMO processing become even more pronounced in certain use cases related to security and surveillance applications, where the concurrent transmission of multiple video streams is often of low rate (e.g., from infrared, thermal, and night-vision devices). Even if the cumulative transmission rate of the transmitted streams remains well below the channel capacity limits, it becomes impossible with conventional MIMO processing to reliably demultiplex more concurrently transmitted streams than AP antennas.
It is also important to notice that in such scenarios, increasing the number of antennas may not be a viable solution.
Besides the increased power demands, additional antennas increase the weight and physical dimensions of the receiving device, reducing its portability and potentially its concealability, key considerations for speed-critical, security-focused applications.
As we show in this demo, departing from conventional processing approaches allows us, with just a single receive antenna, to reliably support up to four live, low-rate video streams concurrently on the same frequency and time resources.

\begin{figure*}[t]
   \includegraphics[]{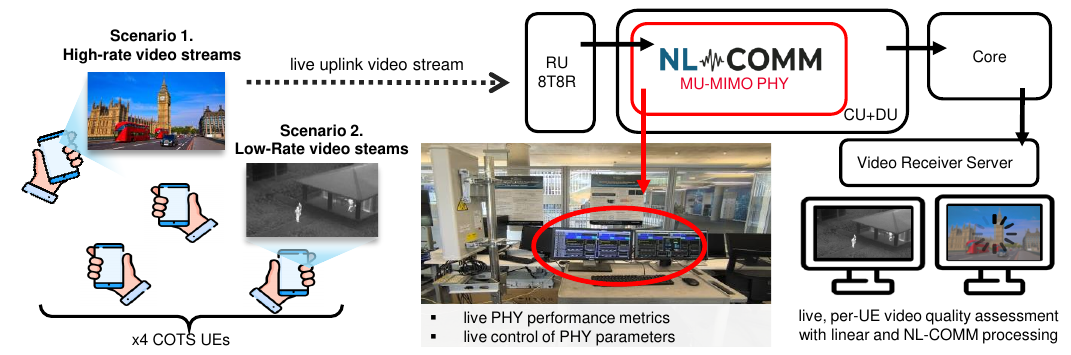}
    \caption{Demonstration setup comparing live stream quality from four concurrently transmitting COTS UEs over our Open-RAN and 5G-NR MIMO PHY system, comparing NL-COMM processing side-by-side with various state-of-the-art linear and nonlinear MIMO detection algorithms.}
    \label{fig:demosetup}
\end{figure*}

An alternative processing approach that accounts for the MIMO channel characteristics,  without degrading its inherent capabilities, is non-linear (NL) MIMO processing which focuses on jointly processing the mutually interfering MIMO streams. Even though non-linear processing approaches \cite{softoutputSD,SFSD,nikitopoulos_geosphere_2014} promise substantial throughput and connectivity gains, their exponentially scaling computational complexity with the number of concurrently supported MIMO streams
and their weak parallelization properties render them unsuitable for real-time realizations, especially in heavily software-based systems (e.g., Open RAN)~\cite{softiphy,azariah2022survey}).

In this demo, we present the latest capabilities of our advanced NL processing framework, NL-COMM \cite{nlcomm2024}. This is the first time non-linear MIMO processing will be demonstrated with commercial off-the-shelf (COTS) user equipment (UEs) in a fully 3GPP-compliant environment. Although NL-COMM’s algorithmic gains have been documented in literature through both simulations and over-the-air experiments \cite{nikitopoulos2024towards,jayawardena_g_multisphere_2020,nikitopoulos_massively_2022,katsaros_vehicular}, this is the first time where these gains will be directly observed as practical improvements at the application layer, allowing the audience of the demo to assess the gains of NL-COMM by directly comparing live video stream transmission of current state-of-the-art MIMO algorithms and NL-COMM.
NL-COMM is a near-product MIMO PHY solution that meets 5G-NR real-time latency requirements entirely in software. This demo also represents the first real-time, over-the-air demonstration of NL-COMM’s extreme connectivity capabilities, supporting four concurrent video streams with a single base station antenna. 
We will showcase NL-COMM’s performance in direct comparison with linear processing methods (i.e., ZF, MMSE) and the most popular non-linear approaches (i.e., successive interference cancellation (SIC)), highlighting the tangible application-layer gains it delivers.

\section{Demonstration Setup}
The demonstration setup, shown in Fig. \ref{fig:demosetup}, consists of two main video streaming scenarios. First, we showcase NL-COMM’s ability to reliably support in real-time and over-the-air high-rate video transmission from four COTS UEs on the same frequency and time resources, using only half the antennas required by linear methods. The second scenario shows the overloading gains of NL-COMM, supporting up to four low-rate video streams (e.g., for security-focused applications) with just a single base station antenna.
This demo utilizes our Open-RAN 5G-NR Stand Alone (SA) system, which supports MU-MIMO with both linear and nonlinear processing. The system is based on a significantly extended and optimized OpenAirInterface software platform. The setup includes a 5G-NR base station with a commercial 8-port antenna array, hosted on a DELL R740 server, and an Ettus USRP X440 software-defined radio, all connected to a 5G Core Network on a separate DELL R740 server, along with four Nokia XR20 UEs.
Attendees will be allowed to adjust the placement of the UEs and, through a specially designed user interface, remotely modify various system parameters. These include the MIMO detection algorithm, the number of supported MIMO streams, and the active antenna elements, allowing them to observe the performance gains achieved by NL-COMM in real-time.
The effects of these adjustments on system performance will be shown via a live video stream, presenting both application-level video quality from each UE and various low-level PHY performance metrics.
Our demonstration setup requires space for three 22-inch computer screens and a high-speed internet connection for live streaming. 

\section*{Acknowledgments}
This work has been supported by the ``NL-COMM'' \cite{nlcomm2024} project, a winner of the SBRI competition by Innovate UK and several UK’s DSIT projects.

\bibliographystyle{IEEEtran}
\bibliography{bib_demo_trunc}

\end{document}